\documentclass[conference]{IEEEtran}

\IEEEoverridecommandlockouts

\usepackage{graphicx}
\usepackage{todonotes}
\usepackage{xurl}
\usepackage{listings}
\usepackage{subcaption}
\usepackage{hyperref}
\usepackage{amssymb}
\usepackage[hyphenbreaks]{breakurl}

\usepackage{multirow}

\usepackage{pgf-umlsd}

\usepackage{tikz}
\usetikzlibrary{arrows.meta}

\lstdefinelanguage{Solidity}{
  keywords={contract, function, modifier, event, enum, struct, mapping, address, uint, bool, string, require, returns, public, private, internal, external, view, pure, payable, new, emit},
  keywordstyle=\color{blue}\bfseries,
  sensitive=true,
  morecomment=[l]{//},
  morecomment=[s]{/*}{*/},
  morestring=[b]",
  morestring=[b]',
  commentstyle=\color{green!50!black},
  stringstyle=\color{red},
  identifierstyle=\color{black},
  numbers=left,
  numberstyle=\tiny\color{gray},
  stepnumber=1,
  numbersep=10pt,
  backgroundcolor=\color{white},
  tabsize=2,
  showspaces=false,
  showstringspaces=false,
  breaklines=true
}

\AtBeginDocument{%
  \providecommand\BibTeX{{%
    \normalfont B\kern-0.5em{\scshape i\kern-0.25em b}\kern-0.8em\TeX}}}

\usepackage[letterpaper, left=1in, right=1in, bottom=1in, top=0.75in]{geometry}

\author{
\IEEEauthorblockN{Valerian Callens}
\IEEEauthorblockA{\textit{Quantstamp, Inc.}\\
valerian@quantstamp.com}
\and
\IEEEauthorblockN{Zeeshan Meghji}
\IEEEauthorblockA{\textit{Quantstamp, Inc.}\\
zeeshan@quantstamp.com}
\and
\IEEEauthorblockN{Jan Gorzny}
\IEEEauthorblockA{\textit{Zircuit}\\
jan@zircuit.com}
}

\begin{document}

\title{Temporarily Restricting Solidity Smart Contract Interactions}

\maketitle

\begin{abstract}

In this work we explore ways to restrict the ability to call Solidity smart contract functions for a specified duration.
We describe methods to restrict functions from being called twice in the same transaction, block, or time period.
This is related to the notion of non-reentrant functions, which are functions that can be called within a previous execution. 
These methods can be used to restrict interactions with entire sets of functions of smart contracts.
We are motivated to revisit this topic for two reasons.
First, we note that sixteen real-world smart contracts exploits in 2023 resulting in over \$136M USD lost or stolen that could have been prevented by restricting function calls.
As part of this survey, we dissect a new class of exploit that involves so-called \emph{read-only} reentrancy: exploits that re-enter read-only functions to make smart contract state inconsistent in order to enable their exploitation.
Second, while some of these approaches are simple, they may not always behave the same across different blockchains that support Solidity.
\end{abstract}

\sloppy
\section{Introduction}\label{sec:intro}
Modern blockchains like Ethereum \cite{ethereum} support so-called \emph{smart contracts}.
A smart contract is a program deployed on the blockchain whose functions can be called via blockchain transactions \cite{szabo1997idea}.
Smart contracts provide the ability to build ``decentralized Applications'' (``dApps'') which use the blockchain as a back-end for logic and data storage.
dApps can hold or implement digital assets such as Ether or ERC-20 tokens \cite{ERC-20}.

A function is reentrant if its body can be re-entered before its previous execution completes.
Reentrant functions have been a source of headaches for the Ethereum community, as their misuse has resulted in stolen and lost cryptocurrencies. 
While reentrancy itself is not a bug but a feature of the dominant smart contract language on Ethereum, Solidity \cite{solidity}, it can in specific conditions have a negative impact on a system \cite{DBLP:conf/isola/DemeyerRV22}.
Perhaps most famously, a reentrant function caused the DAO hack \cite{daohack}, a loss so significant that the Ethereum community was split on how to resolve the issue.

As a result of these problematic issues, many developers choose to prevent their functions from being reentrant altogether, and more generally to conditionally lock some parts of their systems.
The ability to pause a contract can help in some cases, but often cannot be triggered in time.
This ability must typically be called manually, and is impossible to trigger between actions of an arbitrary transaction interacting with a smart contract.
One alternative to manual pausing is to prevent actors from repeatedly interacting with a smart contract in a short period of time.

This paper aims to study ways to restrict repeated access to smart contract functions.
After reviewing common reentrancy prevention methods, we show similar approaches can be used to restrict functions for a particular duration of time (as indicated by the blocks of the underlying blockchain), number of passing blocks, and even from being called twice within the same transaction.
We note that these approaches can be generalized so that the appropriate Solidity function modifiers can be used to restrict (non-singleton) set of functions at the same time, thereby allowing access restrictions to entire contract or dApp logic.
While these techniques may not appear involved, we think that it is important to revisit them at this time for three reasons: they have different properties, they may not behave as expected in modern settings, and they are surprisingly effective.

First, these techniques do not all fit the same use cases.
For example, some may be used to handle related concerns, like so-called \emph{flash loans} (see e.g., \cite{DBLP:conf/fc/QinZLG21}),  while others will not.
Moreover, they may come at the cost of decreased composability, a distinctive property of the Decentralized Finance (``DeFi'') ecosystem~\cite{DBLP:conf/fc/TolmachLLL21}~\cite{DBLP:conf/aft/Werner0GKHK22}.
Additionally, the more complex the lock protection is (e.g., the number of external calls or storage variables used), the more gas (the unit used to pay for and limit computation on Ethereum) it will require and the more it will restrict external interactions.

Second, these approaches may not always behave the same way even though they are written in the same language.
In the past Solidity smart contracts were written almost exclusively for Ethereum; now, there are other chains supporting the language.
Many rollups and other layer two solutions (see e.g., \cite{DBLP:journals/access/ThibaultSH22} \cite{DBLP:conf/fc/GudgeonMRMG20}) use a virtual machine that aims to be similar to the Ethereum Virtual Machine (EVM), but actually differs slightly. %\cite{DBLP:journals/jnca/GangwalGT23}
These differences may include changes to the \texttt{block.timestamp} and \texttt{block.number} operation codes (``opcodes'').
Using these opcodes, or others, may result in protections that are not effective or assumptions that do not hold, when contracts are redeployed from one EVM-based blockchain to another.
It is important to highlight and understand these differences before they cause issues for smart contracts using these methods on such chains.

Finally, these approaches are effective.
In 2023 alone, over \$136M USD was lost or stolen through exploits that spanned only a few transactions or blocks.
Table \ref{tab:hacks} shows the DeFi projects affected along with their estimated losses at the time of the exploit. 
The last four columns indicate which types of restriction suggested in this paper would have been sufficient to prevent the attack.
It is clear from the table that the techniques in this paper may be worth using, even if they break composability with other dApps in some settings.
Moreover, these exploits show that reentrancy is not a problem of the past nor are they limited to a single blockchain. 
Despite being found in EVM bytecode contracts, these issues appeared across a wide range of blockchains: Arbitrum, Avalanche, Binance Smart Chain, zkSync, and  Ethereum itself.
For dApp developers, it is clear that a trade-off must be made between 1) the security of the protocol, 2) the fit between business requirements (including deployment chain), usability and chosen lock protections, and 3) the underlying gas costs.

This paper is organized as follows.
Section \ref{sec:reentrancy-overview} introduces reentrancy and standard mitigation techniques.
Section \ref{sec:prev-reentrancy} reviews standard reentrancy prevention methods. 
Section \ref{sec:access} discusses how sets of functions can be restricted from reentrancy, using a deep dive of a so-called \emph{read-only reentrancy} attack as a motivating example.
In Section \ref{sec:additional-restrictions}, we describe additional techniques for temporarily restricting contract access.
%Section \ref{sec:reentrancy} describes standard techniques for restricting reentrant calls, which we will build upon in the following sections and provides additional context for this work.
These techniques build upon those for standard reentrancy but are more restrictive.
Section \ref{sec:tx} describes a novel approach for preventing functions from being called twice in the same transaction (whether or not it is a reentrant call), and Section \ref{sec:time} takes this a step further to avoid two calls within a fixed duration.
The approaches in these subsections get subsequently more restrictive but are not without their own merits.
%Section \ref{sec:access} illustrates how the approaches of the previous three subsections can be applied to an entire dApp, rather than just a single transaction.
%In each of these subsections, the incidents that were preventable by that section's technique, as well as the trade-offs for the technique, are presented.
Along the way, we describe how a restriction could have prevented or mitigated the attacks in Table \ref{tab:hacks}.
Note that some incidents may have been prevented by multiple of these approaches, but we describe them in the context of the least restrictive mitigation strategy.
%; we attempt to discuss each incident in the section most closely related to the corrective measure chosen by the project's team after the incident occurred.
%the incidents will be described so that an appropriate prevention method can be suggested to avoid repeated issues.
%Within 
Finally, we review related work in Section \ref{sec:related} before concluding the paper in Section \ref{sec:conclusion}.

\begin{table}[t]
    \centering
    \begin{tabular}{|l|c|c|c|c|c||c c|}
    \hline %\rotatebox[origin=c]{90}{Loss Amount}
         Project &  Loss & \multicolumn{6}{c|}{Restriction}  \\  \cline{3-8}
          & & \multicolumn{2}{c|}{\rotatebox[origin=c]{90}{Some Blocks } \rotatebox[origin=c]{90}{or Time}} & \rotatebox[origin=c]{90}{Same Transaction} & \rotatebox[origin=c]{90}{Same Execution} &  \multicolumn{2}{c|}{\rotatebox[origin=c]{90}{Sets of} \rotatebox[origin=c]{90}{Functions}  }   \\ 
        \hline \hline
        Conic Finance \cite{conic} & \$3.6M &  \multicolumn{2}{c|}{\textcolor{gray}{\checkmark}} & \textcolor{gray}{\checkmark}  & {\checkmark}  & \multicolumn{2}{c|}{\checkmark} \\ \hline
        Curve \cite{curve} & \$73.5M &   \multicolumn{2}{c|}{\textcolor{gray}{\checkmark}}  &  \textcolor{gray}{\checkmark} &  \checkmark$^*$ & &\\ \hline
        dForce \cite{dforce}  & \$3.65M & \multicolumn{2}{c|}{\textcolor{gray}{\checkmark}}  & \textcolor{gray}{\checkmark} &  \checkmark & \multicolumn{2}{c|}{\checkmark}\\ \hline
        EraLend \cite{eralend}  & \$3.4M &  \multicolumn{2}{c|}{\textcolor{gray}{\checkmark}}  & \textcolor{gray}{\checkmark} & \checkmark & \multicolumn{2}{c|}{\checkmark} \\ \hline
        Exactly \cite{exactly} & \$7.3M & \multicolumn{2}{c|}{\textcolor{gray}{\checkmark}}  & \textcolor{gray}{\checkmark} & \checkmark && \\ \hline
        Hundred \cite{hundred}  & \$7M &   \multicolumn{2}{c|}{\textcolor{gray}{\checkmark}} & \checkmark &  & &\\ \hline
        LendHub \cite{lendhub} & \$6M &    \multicolumn{2}{c|}{\textcolor{gray}{\checkmark}} & \checkmark &  & \multicolumn{2}{c|}{\checkmark} \\ \hline
        Midas \cite{midas}  & \$660K &  \multicolumn{2}{c|}{\textcolor{gray}{\checkmark}} & \textcolor{gray}{\checkmark} & \checkmark & \multicolumn{2}{c|}{\checkmark}\\ \hline
        Orion \cite{orion}  & \$3M &  \multicolumn{2}{c|}{\textcolor{gray}{\checkmark}} & \textcolor{gray}{\checkmark} & \checkmark & & \\ \hline
        Palmswap \cite{palmswap}  & \$900K &   \multicolumn{2}{c|}{\textcolor{gray}{\checkmark}} & \textcolor{gray}{\checkmark} & \checkmark & &\\ \hline
        Platypus (Feb) \cite{platypus1} & \$8.5M &  \multicolumn{2}{c|}{\textcolor{gray}{\checkmark}} &\checkmark  &   &  \multicolumn{2}{c|}{\checkmark}\\ \hline
        Platypus (Oct) \cite{platypus2}  & \$2.2M &  \multicolumn{2}{c|}{\checkmark} &  &  &  \multicolumn{2}{c|}{\checkmark}\\ \hline
        Sentiment \cite{sentiment} & \$1M &  \multicolumn{2}{c|}{\textcolor{gray}{\checkmark}}& \textcolor{gray}{\checkmark} & \checkmark & \multicolumn{2}{c|}{\checkmark}\\ \hline
        Stars Arena \cite{starsarena} & \$3M  &  \multicolumn{2}{c|}{ \textcolor{gray}{\checkmark} } &  \textcolor{gray}{\checkmark}  & \checkmark & &\\ \hline
        Sturdy \cite{sturdy} & \$800K &  \multicolumn{2}{c|}{\textcolor{gray}{\checkmark}} & \textcolor{gray}{\checkmark} & \checkmark & \multicolumn{2}{c|}{\checkmark}\\ \hline
        Yearn \cite{yearn} & \$11M &  \multicolumn{2}{c|}{\textcolor{gray}{\checkmark}} & \checkmark & &  \multicolumn{2}{c|}{\checkmark} \\ \hline
    \end{tabular}
    \caption{A list of DeFi projects exploited in 2023 with their estimated losses. These projects were exploited in ways that could have been prevented by temporarily restricting the attacker's ability to access the relevant smart contract functions.
    %Every third column indicates the method in this paper suggested to prevent the incident from occurring. 
    Columns three to six indicate the methods described in this paper that could have prevented the incident from occurring, with the least restrictive approach in bold. 
    A checkmark in column six indicates that a restriction should be applied to several functions to be effective.
    Accessing smart contract functions can be restricted to avoid calls within the same function execution (reentrancy; Section \ref{sec:prev-reentrancy}), within the same transaction (Section~\ref{sec:tx}), and within some number of blocks or a time period (Section~\ref{sec:time}). Access restrictions for sets of functions are discussed in Section~\ref{sec:access}. Curve's suggestion is marked with a $^*$ because it did have such a guard, but a compiler bug rendered it ineffective (also explained in Section~\ref{sec:prev-reentrancy}).}
    \label{tab:hacks}
\end{table}

\section{Reentrancy}\label{sec:reentrancy-overview}

In this section we first review reentrancy (Section \ref{sec:prev-reentrancy}) and its standard mitigation strategies.
Then we explore how these strategies can be applied to non-singleton sets of functions in order to guard entire contracts (and therefore dApps) from being reentered through different functions (Section \ref{sec:access}).

\subsection{Restricting calls within the same execution (reentrancy)}\label{sec:prev-reentrancy}

A function is said to be \emph{reentrant} if it can be called before completing its previous execution.
A dangerous reentrant function often involves an external function call to another smart contract, which can be manipulated.
This external function call may pass control of the execution to the called smart contract, which may be arbitrary (and therefore malicious).
In turn, these functions may reenter the original function, or another function on the contract, to exploit a vulnerability.
In this section, we describe three common patterns for limiting reentrancy: a design pattern, gas limiting, and a function modifier.

\subsubsection{Checks-Effects-Interactions}\label{sec:checks}

One way to mitigate potential reentrancy is to design functions so that the reentrancy does not have an effect, even if it occurs.
This can be achieved by using the so-called \emph{Checks-Effects-Interactions} design pattern~\cite{ma2019fundamentals}.
This design pattern dictates that all pre-conditions of the function are checked first (checks), then state updates are recorded (effects), and finally other addresses are called (interactions). 
This design pattern does not introduce additional storage overhead for mitigating reentrancy, but may be difficult to implement when there are complex checks, effects, or interactions.

\subsubsection{Gas limiting external calls}\label{sec:gas-limit}

Recall that external calls are a source of dangerous reentrancy. 
The external call can be the execution of a smart contract function conforming to an interface, or a low-level instruction. 
A low-level instruction may be used to send native currency (Ether) destination address.
This destination address may be a contract that is able to receive this currency, meaning that it implements either a \texttt{payable} \texttt{fallback()} or \texttt{receive()} function, which are designed for this use case. 
%%%For the last two options, an amount of gas from the remaining amount of gas at this step of the execution is forwarded to the address that will be called, for execution purposes. 
There are three low-level instructions to send Ether: \texttt{transfer()}, \texttt{send()}, and \texttt{call()}. 
It is possible to limit that amount of gas forwarded, which in turn limits the number of actions that can be executed within that call, ultimately reducing the likelihood of a reentrancy attack during that call. 
The instructions \texttt{transfer()} and \texttt{send()} only forward an hard-coded amount of \texttt{2300} gas, which makes reentrancy unlikely.
However, \texttt{call()} does not limit that amount and by default, the remaining transaction gas is forwarded to the address that will receive the Ether, thereby possibly triggering a function call.
This call may take more (malicious) actions.
To avoid this, the developer must explicitly limit the amount of forwarded gas in this call. %(ref: https://solidity-by-example.org/sending-ether/) 
Note that the behavior of these low-level instructions, and in particular the amount of forwarded gas, may differ in future updates to the EVM or on other EVM-based blockchains.

\subsubsection{Nonreentrant Modifier}\label{sec:standard}
A simple implementation to protect a function from reentrancy was proposed by OpenZeppelin \cite{ozreentrancy}. 
%Figure \ref{fig:oz} shows a simplified version of their implementation in Solidity (v4.5.0).
In order to add a reentrancy guard to a contract, the contract must inherit from the contract \texttt{ReentrancyGuard} and add the modifier \texttt{nonReentrant} to any function that should be protected. 
This is a way to make sure there are no reentrant calls to the set of protected functions (which may be a singleton). 
The modifier implementation consists of one state variable: the mutual exclusion variable, or ``mutex''. 
Before the execution of a modified function, the function will revert if the mutex already has the locked state. %(line 9) 
If this first check passes, the state of the mutex is modified to represent the locked state. %(line 10) 
After that, the execution of the function is triggered. %(line 11) 
Once the function is executed, the value of the mutex is set back to the unlocked state. %(line 12)
This approach is simple and effective, with minimal additional gas overhead.

\subsubsection{Preventable Incidents}\label{sec:prev-inc-reent}

The Exactly Protocol incident \cite{exactly} was executed in multiple steps, involving reentrancy.
First, the attacker was able to bypass a \texttt{permit} check\footnote{Some ERC-20 tokens implement EIP-2612 \cite{erc20permit} which allows users to sign a message off-chain (the ``permit'') indicating that another user can spend these tokens; the signed permits are checked on-chain.} to trick the protocol into operating on behalf of the \texttt{permit} issuer rather than the true caller, and then reenters a function on behalf of the \texttt{permit} issuer.
While a core failure of the system was the ability to fool the protocol into behaving on behalf of the \texttt{permit} issuer, a \texttt{nonReentrant} modifier on the appropriate function may have mitigated the damage.
This type of impersonation for \texttt{permit} issuers has been previously exploited for cross-chain bridges \cite{DBLP:conf/icbc2/LeeMDG23}.

%%% Exactly
% https://www.immunebytes.com/blog/defi-exactly-protocol-hack-analysis/
% https://twitter.com/De_FiSecurity/status/1692494879580991528/photo/1

%%% Curve
% https://www.theblock.co/post/244438/curve-finance-has-recovered-70-of-hacked-funds-with-distribution-planned

A failure to mark a deposit function as \texttt{nonReentrant} was the core problem of the Orion Protocol incident~\cite{orion}.
An attacker was able to deposit a token with a malicious contract, which when transferred, reentered Orion protocol.
Using this alongside real stablecoins (see e.g., \cite{DBLP:journals/cacm/ClarkDM20}), they were able to fool the protocol that they were depositing more than than they actually did, and then withdrew that entire amount.
The Stars Arena incident~\cite{starsarena} was similar, though it targeted AVAX tokens instead on the Avalanche chain.

The Curve Finance incident \cite{curve} could have been prevented by a functional \texttt{nonReentrant} modifier on the \texttt{addLiquidity()} function.
In fact, the function had a similar guard on it, but the contracts were written in the Vyper (see e.g., \cite{DBLP:conf/brains/KaleemML20}) smart contract language, whose compiler contained a bug that affected the implementation of this guard.
This incident highlights the need to consider other guard options and to ensure that the entire stack can be trusted.

\subsubsection{Trade-Offs}\label{sec:tradeoffs-reent}
The \texttt{nonReentrant} modifier discussed in this section trades optimal gas usage for extra security.
The modifier's extra state variable is minimal and unavoidable but greatly reduces the complexity of the system by preventing reentrancy all together for the modified functions.
This does not affect the system's composability unless dependent dApps are expecting to reenter the function, which can likely be factored out.
While both the modifier and the checks-effects-interactions pattern are unlikely to have different semantics on different blockchains, limiting gas may be problematic if gas usage differs over time or on different chains.
The trade-offs for gas limiting is therefore less clear: there is no guarantee that the extra security will remain as the dApp is redeployed or over time.
Moreover, all of these approaches do not guard against exploits which span a larger duration; such exploits would indeed be non-reentrant but could be problematic nonetheless.

\subsection{Restricting sets of function calls}\label{sec:access}

Restricting function access is typically considered at the contract level aiming to prevent a modified functions from being reentrant.
However, it can also be considered at a system level and to restrict arbitrary subsets of functions simultaneously.

%\subsection{Fine-grained access restrictions}

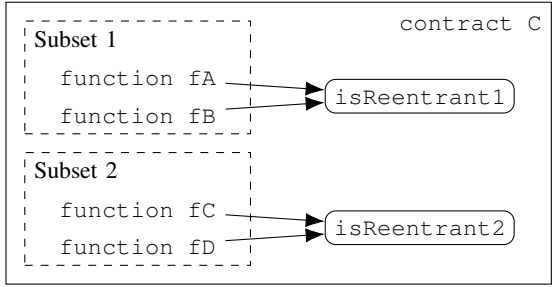
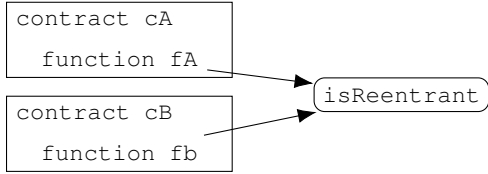
\begin{figure}[bt]
    \centering
    \begin{subfigure}{0.45\textwidth}
    \centering
    \begin{tikzpicture}[scale=1]
    % Outer box
    \draw (0,2.25) rectangle (7.25,6);
    \node [anchor=south east] at (7.25,5.5) {\small \texttt{contract C}};
    
    \draw[dashed] (0.25,2.5) rectangle (3.25,4);
    \node [anchor=north west] at (0.25,4) {\small Subset 2};

    \draw[dashed] (0.25,4.25) rectangle (3.25,5.75);
    \node [anchor=north west] at (0.25,5.75) {\small Subset 1};

    \node[draw, rounded corners] (subset1) at (5.5,4.75) {\small \texttt{isReentrant1}};
    \node (fa) at (1.75,5) {\small\texttt{function fA}};
    \draw[-{Latex[scale=1.5]}] (fa) -0 (subset1);
    \node (fb) at (1.75,4.5) {\small\texttt{function fB}};
    \draw[-{Latex[scale=1.5]}] (fb) -0 (subset1);

    \node[draw, rounded corners] (subset2) at (5.5,3) {\small\texttt{isReentrant2}};
    \node (fc) at (1.75,3.25) {\small\texttt{function fC}};
    \draw[-{Latex[scale=1.5]}] (fc) -0 (subset2);
    \node (fd) at (1.75,2.75) {\small\texttt{function fD}};
    \draw[-{Latex[scale=1.5]}] (fd) -0 (subset2);
    
    %%%%\node [anchor=south west] at (3.5,1) {Subset 1};
\end{tikzpicture}

    \caption{The contract \texttt{C} has two subsets of functions, $\{$\texttt{fA}, \texttt{fB}$\}$ and $\{$\texttt{fC}, \texttt{fD}$\}$. Access is restricted at the level of each subset using distinct mutexes.}
    \label{fig:subset-level}     
    \end{subfigure}
    \hfill
    \begin{subfigure}{0.45\textwidth}
    \centering
    %\vfill %%%jgorzny: useless!
    %\includegraphics[width=\textwidth]{figures/Reentrancy-system.png}
        \begin{tikzpicture}[scale=1]
    % Outer box

    \draw (0.25,3.5) rectangle (3.25,4.5);
    \node [anchor=north west] at (0.25,4.5) {\small\texttt{contract cB}};

    \draw(0.25,4.75) rectangle (3.25,5.75);
    \node [anchor=north west] at (0.25,5.75) {\small\texttt{contract cA}};

    \node[draw, rounded corners] (subset1) at (5.5,4.52) {\small\texttt{isReentrant}};
    \node (fa) at (1.75,5) {\small\texttt{function fA}};
    \draw[-{Latex[scale=1.5]}] (fa) -0 (subset1);

    \node (fb) at (1.75,3.75) {\small\texttt{function fb}};
    \draw[-{Latex[scale=1.5]}] (fb) -0 (subset1);
    
    %%%%\node [anchor=south west] at (3.5,1) {Subset 1};
\end{tikzpicture}
    \caption{A system composed of two contracts, \texttt{cA}, which defines a function \texttt{fA}, and \texttt{cB}, which defines a function \texttt{fB}. Access control is enforced at the level of the system, using a common state variable as mutex that is part of all system contracts.}
    \label{fig:system-level}  
    \end{subfigure}
    \caption{Preventing smart contract access through shared variables.}
    \label{fig:slvels} 
\end{figure}

The OpenZeppelin non-reentrant guard (Section \ref{sec:prev-reentrancy}) is at the contract level, since it requires writing to a variable that is outside the scope of the function to which the modifier is applied.
If this modifier is applied to more than one function (as is often the case), the approach is already used to restrict access to that set of functions.
%The simplest implementation of a reentrancy protection consists (Figure \ref{fig:oz}) of a unique state variable which mutually excludes the execution of a set of functions in nested calls.
However, this solution is not adapted to the case where distinct subsets of functions within the same contract should be protected by a distinct reentrancy protections, such that functions from one subset should still be able to call a function of the other subset within a nested call, and vice-versa. 
In this case, one dedicated state variable should be defined per subset of functions. 
This is illustrated in Figure \ref{fig:subset-level}.

%\subsection{System level restrictions}
It is also possible to enforce system-wide access restrictions to prevent any nested call triggered by a set of functions.
For example, Figure \ref{fig:system-level} shows two contracts \texttt{cA} and \texttt{cB} which implement functions \texttt{fA} and \texttt{fB}, respectively. 
By using a shared mutex variable common to the guard modifiers on functions \texttt{fA} and \texttt{fB}, one can make it impossible to \texttt{fA} while \texttt{fB} is executing, or vice versa. 
If the state variable is not on the same contract as the associated functions, a cross-contract access control mechanism should be enforced to restrict the list of addresses allowed to update this shared variable.

This is particularly helpful in DeFi settings.
For example, one could use the same duration-based mutex variable for a \texttt{deposit()} and \texttt{withdraw()} function for lending dApps. 
In such a design, a user could deposit as much as they want within any duration, but would only be able to withdraw after a delay.
This can mitigate flash loan concerns without negatively affecting many dApp users.

\subsubsection{Read-only Reentrancy}
\emph{Read-only reentrancy} is related to function reentrancy, and occurs when a read-only (i.e.~\texttt{view}) function $f'$ is entered while another function of the same dApp is executing $f$.
Notably, the read-only function $f'$ being entered may not be the same one being executed by the dApp ($f$), and that may be problematic as $f$ may be updating values that $f'$ is reading.
In such a case, the function $f'$ does not return accurate values; it may report values inconsistent with the state of the dApp both before and after the execution of $f$ began.
Any contract that relies on the accuracy of this value will then be subject to exploitation.
dApps may prevent read-only reentrancy by adding a check before calling the read-only function to ensure the smart contract is not being reentered, but this involves making the function modify state (increasing the gas cost of these otherwise cheap \texttt{view} functions).

\subsubsection{Sentiment Incident Deep-Dive}\label{sec:sentimenet}

%https://hackmd.io/@sentimentxyz/SJCySo1z2 %%best
%https://blog.solidityscan.com/sentiment-hack-analysis-reentrancy-attack-8d1b2b6a1691
%https://twitter.com/CertiKAlert/status/1643364285182013441
%https://medium.com/coinmonks/theoretical-practical-balancer-and-read-only-reentrancy-part-1-d6a21792066c %%% least relevant

\begin{figure}
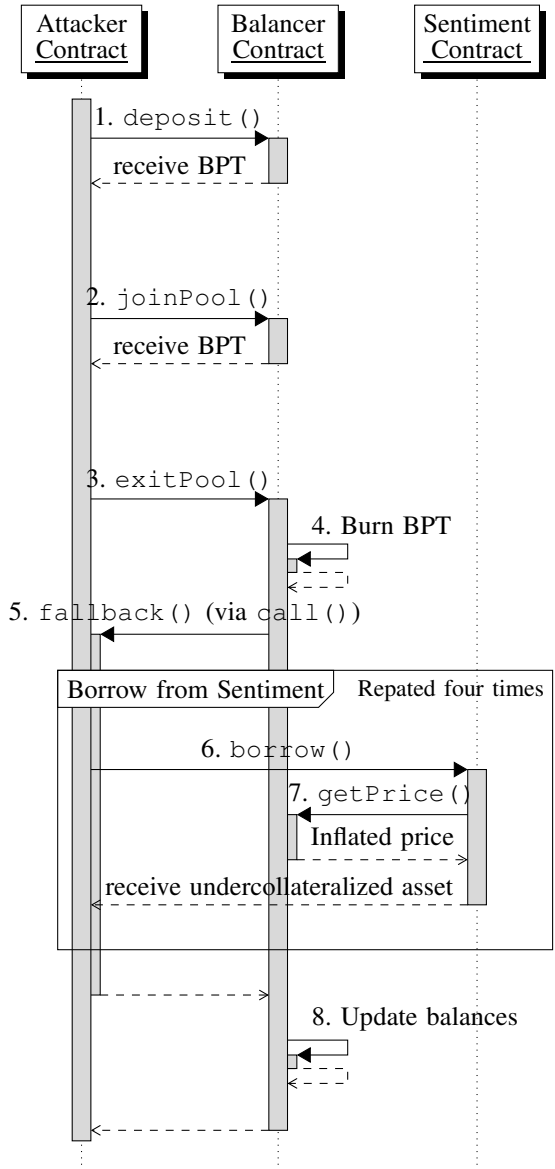

    \begin{sequencediagram} 
    \newthread{a}{\shortstack{Attacker\\Contract}} 
    \newinst[1]{b}{\shortstack{Balancer\\Contract}} 
    \newinst[1]{s}{\shortstack{Sentiment\\Contract}} 
    %\newinst[1]{i}{Attacker Contract} 

    %Step 1
    %%%%\begin{sdblock}{Step 1}{\shortstack{Special Task\\ Controller}} 
    %\begin{sdblock}{Step 1}{} 
        \begin{call}{a}{1. \texttt{deposit()}}{b}{receive BPT} %Line 77
        \end{call} 
    %\end{sdblock} 
    
    \postlevel
    \postlevel
    
    %Step 2
    %\begin{sdblock}{Step 2}{} 
        \begin{call}{a}{\shortstack{2. \texttt{joinPool()}}}{b}{receive BPT} % Line 196
        \end{call} 
    %\end{sdblock}     

    \postlevel
    \postlevel
    %Step 3
    %\begin{sdblock}{Step 3}{} 
        \begin{call}{a}{\shortstack{3. \texttt{exitPool()}}}{b}{}  %Line 250
            \begin{call}{b}{4. Burn BPT}{b}{} 
            \end{call} 
            \setthreadbias{east}
            \begin{call}{b}{5. \texttt{fallback()} (via \texttt{call()})}{a}{} %Line 262
                %\setthreadbias{center}
                
                \begin{sdblock}{Borrow from Sentiment}{Repated four times}  %This call is repeated three additional times.
                \setthreadbias{center}
                \begin{call}{a}{6. \texttt{borrow()}}{s}{receive undercollateralized asset} % Line 263
                    \setthreadbias{east}
                    \begin{call}{s}{7. \texttt{getPrice()}}{b}{Inflated price} 
                    \end{call} 
                    \setthreadbias{center}
                \end{call} 
                
                \end{sdblock} 
                
                \setthreadbias{east}
            \end{call}    
            \setthreadbias{center}
            \begin{call}{b}{8. Update balances}{b}{} 
            \end{call} 
        \end{call} 
    %\end{sdblock}         
    \end{sequencediagram}
    
    \caption{The Sentiment incident, illustrated. The entire exploit takes place during a single transaction on Arbitrum. The \texttt{view} function \texttt{getPrice} is entered more than once during this duration and is used to compute values when an invariant does not hold \emph{within} another execution being performed by the Balancer contract.}
    \label{fig:sentiment-explained}
\end{figure}

Sentiment suffered a \$1M incident which related to to read-only reentrancy \cite{sentiment}.
The Sentiment exploit involves in several steps.
It is illustrated in Figure \ref{fig:sentiment-explained} and described below following the team's own post-mortem \cite{sentiment-pm}.

\noindent
{\bf Step 0.} The attacker acquires a large amount of WETH (Wrapped Ether, which is an ERC-20 representation of Ether) \cite{weth}, WBTC (Wrapped Bitcoin) \cite{wbtc}, and USDC (a USD backed stablecoin) \cite{usdc}. 
    In the actual attack, this is via a flash loan (see e.g. \cite{DBLP:conf/fc/QinZLG21}), where they acquire about $10,050$ WETH, $606$ WBTC, and $18,000,000$ USDC. 
    We omit the details of how the flash loan was executed due to page constraints.
    They use a previously deployed contract that allows the following sequence of events to happen (and all the necessary book keeping and other functionality).

    The \emph{Balancer-33 WETH-33 WBTC-33 USDC} pool of the Balancer protocol \cite{balancer} initially contains $x=616.3996$ WETH, $y=40.9350$ WBTC, and $z=1,155,172.1668$ USDC at the beginning of the exploit. 
    Balancer Pool Token (BPT) is provided by adding liquidity to a pool: the more liquidity one adds, the more BPT they earn (see e.g.,~\cite{DBLP:journals/csur/XuPCF23}).
    The Sentiment protocol allows users to deposit BPT in order to use the BPT as collateral for loans.
    There is a total supply $w = 8412.43882$ of BPT for the pool, each with a price $p=0.220118561$ ETH per token.

\noindent
{\bf Step 1.} The attacker adds WETH to Balancer, in order to gain BPT (transaction 1, \texttt{deposit()}). 
    Their ultimate goal is to manipulate the price of BPT to inflate their collateral value such that Sentiment allows them to borrow funds worth far more than the real value of their collateral.

    The attacker adds $50$ WETH to the \emph{Balancer-33 WETH-33 WBTC-33 USDC}; this increases the amount of WETH in the pool to $x'=616.3996$, the total supply of BPT tokens to $w'=8633.653333$, and price to $p'=0.220127737$.
    That is, there is more WETH in the Balancer contract than initially, the price of BPT is increased slightly, and some BPT tokens are minted for the attacker.

\noindent
{\bf Step 2.} The attacker gains a large value of BPT by adding all of the the pool assets in proportion (transaction 2; \texttt{joinPool()}). 
    This does not materially change the price but grants the attacker about $130,601$ BPT tokens nonetheless; namely, $p''\approx p'$, but $x''=10,666.3996$, $y''=646.9350$, and $z''=19,155,172.1668$ and $w''=139234.634$.

\noindent
{\bf Step 3.} The attacker exits the pool they just deposited into (\texttt{exitPool()}, transaction 3), returning $w''-w'=130,601$ BPT. 
    Critically, they also request that the returned WETH is actually sent as regular ETH.
    The ordering of the effects within the \texttt{exitPool()} is also important: the first thing it does is burn the BPT it receives (transaction 4; this may not be done via a helper function in the actual implementation but it is presented as one for emphasis).
    Next, it returns funds, and finally it updates its internal records (transaction 8; also presented as a function for emphasis).
    
    Recall that this transaction is also executed by the attacker's contract, which also receives the ETH that is sent by Balancer. 
    This contract has a \texttt{payable fallback()} function which is executed when the ETH is sent to the contract via a \texttt{call()} function (transaction 5; c.f.~Section \ref{sec:gas-limit}).
    This allows the attacker to take control and re-enter the state of the Balancer pool at an opportune time to take a large loan against the inflated value of their BPT collateral.
    When the attacker's contract is in control, it calls the borrow function (transaction 6) four times.
    Each time it does this, the \texttt{borrow()} function calls the Sentiment smart contract, which in turn evaluates the attacker's remaining $w'-w=50$ BPT tokens by calling \texttt{getPrice()} on the Balancer contract.
    This means that they are re-entering the Balancer contract while the Balancer contract is still  returning them funds.
    The point of re-entrance is the \texttt{getPrice()} function which uses the total supply of BPT tokens and its internal accounting to calculate the price of the BPT.
    However, since the BPT have been burned but the balances are not yet updated, the calculation is incorrect and the price of the BPT is incorrectly computed as $p'''=3.55007307$ ETH, which is more than sixteen times greater than what it should be.
    Therefore, each time the borrow function is called, more than sixteen time the real value of funds are considered safe loan amounts, and Sentiment allows this value to be taken out as loans.
    The attacker calls \texttt{borrow()} four times, borrowing a total of $81$ WETH, $461,000$ USDC, $361,000$ USDT (another USD stablecoin) \cite{usdt}, and $125,000$ FRAX tokens \cite{frax}.
    Some of these funds are used to pay back the initial flash loan; the rest are withdrawn to the attacker.
    Finally, control is once again passed to Balancer which updates its internal accounting and corrects the price of BPT.
    However, by this time, the attacker has taken a disproportionate amount of funds out of the Sentiment protocol and the damage is done.

\noindent
{\bf Fix.} 
%The Sentiment incident could have been prevented in a number of ways.
The Sentiment team updated their implementations of the function to add a no-op which effectively acts a re-entrancy guard.
However, the issue may have been avoided altogether if the \texttt{exitPool()} function of the Balance protocol followed the Checks-Effects-Interactions pattern (Section \ref{sec:checks}).

%%SIMPLIFIED BELOW

\subsubsection{Preventable Incidents}\label{sec:prev-inc-lvl}

Conic Finance \cite{conic},  dForce \cite{dforce} (on Arbitrum and Optimism), EraLend \cite{eralend} (on zkSync), Midas Capital \cite{midas}, and Sturdy Finance \cite{sturdy} also suffered incidents stemming from problematic read-only reentrancy.

The techniques in this section could have also been used to prevent a number of attacks, like the Yearn Finance incident \cite{yearn}.
In this incident, Yearn suffered from a misconfigured setup rather than reentrancy; coupled with a flash loan, an attacker was able steal funds from the protocol. 
The standard \texttt{nonReentrant} modifier would not have provided protection, however, the duration-based modifiers of Section \ref{sec:time} applied to the relevant deposit and withdrawal functions would have complicated the attack and may allowed Yearn to intervene if it was detected.

%%% Conic
% https://www.immunebytes.com/blog/conic-finance-detailed-hack-analysis-july-21/

%%% dForce
% https://crypto.news/dforce-confirms-the-return-of-exploited-3-65m-to-their-vaults/
% https://crypto.news/dforce-loses-3-65-million-in-a-hack-attack-reports-show/

%%% Midas
% https://medium.com/midas-capital/midas-exploit-post-mortem-1ae266222994

\subsubsection{Trade-Offs}\label{sec:tradeoffs-systemwide}
These approaches trade both developer and user experience for security.
The main difficultly in these approaches is to correctly identify the appropriate subsets of functions to share a mutex variable.
Marking too many functions (especially read-only) functions may make it expensive for users and other dApps to read values from the contract.
Other dApps which try to implement read-only functions that call a guarded read-only function will also be affected, as those functions will no longer be truly read-only.
Therefore, care should be taken to ensure that the smart contract remains usable and composable.
\section{Additional Restrictions}\label{sec:additional-restrictions}

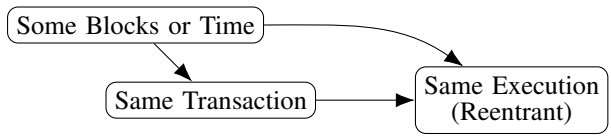
\begin{figure}
    \centering
    \begin{tikzpicture}[scale=1]
    \node[draw, rounded corners] (bt) at (-1,3) {Some Blocks or Time};
    \node[draw, rounded corners] (ro) at (0,2) {Same Transaction};
    \node[draw, rounded corners] (re) at (4,2) {\shortstack{Same Execution\\(Reentrant)}};
    %\node (fc) at (1.75,3.25) {\texttt{function fC}};
    \draw[-{Latex[scale=1.5]}] (bt) -0 (ro);
    \draw[-{Latex[scale=1.5]}] (bt) to [out=0,in=145] (re);
    \draw[-{Latex[scale=1.5]}] (ro) -0 (re);
\end{tikzpicture}
    \caption{Functions which cannot be called twice within some number of blocks (or a given time frame) cannot be done so even if they they do not directly modify function; similarly, such functions cannot be re-entered. Functions which cannot be entered twice in the same transaction also cannot be reentered before their completion.}
    \label{fig:implications}
\end{figure}

In this section, we consider additional techniques for preventing access to functions over different time periods.
Recall that Section \ref{sec:prev-reentrancy} describes a standard function modifier to prevent it all together. %design pattern to nullify the effect of reentrancy in a function
This restricts functions from being re-entered during a previous (ongoing) execution.
Section \ref{sec:tx} describes a technique to restrict a function from being called more than once within the same transaction, which implies that it cannot be called during an existing execution (i.e., it guards against reentrancy too).
Additional modifiers of this kind are presented in Section \ref{sec:time}, which describes how to restrict calls to a function over a period of time or number of blocks.
Clearly, if a function cannot be called twice within some number of blocks, it cannot be called twice in the same transaction (which is contained in a single block) and it cannot be reentrant.
Figure \ref{fig:implications} shows the relationship between these restriction techniques.
%Finally, Section \ref{sec:access} describes use any of these techniques to guard against \emph{sets} of functions (and therefore entire dApps).

\subsection{Restricting calls within the same transaction}\label{sec:tx}

\begin{figure}[bt]
    \centering
\begin{lstlisting}[language=Solidity,firstnumber=1,escapeinside=||]
modifier calledMaxOncePerTransaction() {
    address addressToCheck = address(uint160(bytes20(blockhash(block.number))));
    uint256 initialGas = gasleft();
    uint256 temp = addressToCheck.balance;
    uint256 gasConsumed = initialGas 
        - gasleft();
    require(gasConsumed == 2631,
     "already called in this transaction");
    _;
}
\end{lstlisting}
\caption{A modifier that allows multiple function calls within the same Ethereum block, but no more than once per transaction. Modified functions can only be called once per transaction, leveraging the fact that the gas cost is different when the balance of an address is accessed again within the same transaction. For the sake of simplification, the value \texttt{2631} is hardcoded and includes some  overhead for other necessary operations performed by the modifier.}
\label{fig:txbased}

\end{figure}

One can enforce a transaction-wide lock to make sure that a function can only be called once within the same transaction but multiple times within the same block (i.e., in distinct transactions). 
The implementation of such a protection is challenging because most of the EVM opcodes available provide context at the block level, rather than the transaction level.
Nevertheless, an example is presented in Figure~\ref{fig:txbased}.  
The \texttt{calledMaxOncePerTransaction} modifier presented uses the subtlety of dynamic gas evaluation within the EVM to let a function detect if it is the first time that it is executed within a transaction~\cite{opcodes}. 
In the EVM, if a function tries to get the balance of a given address more than once, a gas discount will happen the second and subsequent accesses.
The first access to the address' balance is said to be \emph{cold} while any subsequent accesses are said to be \emph{warm}.
A cold access costs $2600$ gas, while a warm one costs only $100$ gas.
As a result, it is possible to check via the \texttt{gasleft()} function (which returns the remaining gas for the executing transaction) if the amount of gas used to check such balance is equal or different to the expected cost of accessing this state variable for the first time. 
Note that using a hard coded value for this gas amount could lead to breaking changes if the gas calculation of the network is modified during a later update of the EVM.

There are a few ways the modifier in Figure \ref{fig:txbased} can be improved.
The example uses a pseudo-random address on line~2 to prevent honest users from accidentally preventing their own calls to this function from reverting.
A more secure randomness approach could be beneficial.
Moreover, by updating the \texttt{addressToCheck} variable after the function call (line~9) to another (random) address, this modifier will prevent reentrancy, but allow multiple calls to the modified function in a transaction.
By keeping track of the number of times this address changes, one can increase the maximum number of times the function is called in a single transaction.
Moreover, the address determined on line~2 could be a function of the \texttt{msg.sender} (or \texttt{tx.origin}), in order to make this protection apply to one actor at a time.

\subsubsection{Preventable Incidents}\label{sec:prev-inc-tx}

This approach would have been effective mitigating a number of attacks.
In the Sentiment exploit (Section \ref{sec:sentimenet}) the \texttt{getPrice()} function (although a read-only one) could have been guarded with the modifier in this section, which would have been effective as it was called several times.
Platypus Finance suffered two incidents that could have also have been mitigated.
In the first incident \cite{platypus1}, the attacker was able to take advantage of an ``emergency withdrawal'' functionality after immediately depositing funds to borrow, in order to withdraw without paying back the newly incurred debt.
Flash loan attacks were cause of the LendHub incident \cite{lendhub} and Palmswap incident \cite{palmswap}, both on the Binance Smart Chain, and the Hundred Finance incident \cite{hundred} on Ethereum. 
All of these would benefited from delays between deposits and withdrawals, achieved by placing the guard on both of these functions.
A small delay may protect most honest actors while allowing protocols to intervene if malicious actions are suspected.

%%%% Hundred finance:
% https://www.numencyber.com/hundred-finance-exploit-7-million/
% https://twitter.com/HundredFinance/status/1649881857389916162
% https://blog.hundred.finance/15-04-23-hundred-finance-hack-post-mortem-d895b618cf33

%%%% LendHub:
% https://twitter.com/SlowMist_Team/status/1613906592520409088

%\input{sections/sentiment} %% old location

\subsubsection{Trade-Offs}\label{sec:tradeoffs-tx}
The trade-offs for this approach are less straightforward.
While it may be undesirable to prevent DeFi users from efficiently borrowing several assets in a single transaction, it be desirable to prevent users from making smart contracts which repeatedly mint Non-Fungible Tokens (NFTs; ERC-721 tokens \cite{ERC-721}) during a new release.
In the former example, gas and user experience are traded for security, while in the latter, gas is traded for both user experience and security.
The technique in this section is more application specific than the general reentrancy modifier of Section \ref{sec:prev-reentrancy}.
Regardless, it should be used with caution: it may not be future-proof (due to changes in gas costs) and may behave differently on different chains.

\subsection{Restricting calls within a time period or number of blocks}\label{sec:time}
In the most restrictive setting, functions cannot be called more than once within a particular time period.
This period may be measured by wall-clock time, or by some number of blocks.
Figure \ref{fig:timebased} shows the \texttt{ReentrancyGuardDuration} modifier, a modified version of the OpenZeppelin \texttt{Reentrancy\-Guard} mutex in Section \ref{sec:prev-reentrancy}. %Figure \ref{fig:oz}
In this implementation, the lock is enforced by the passing of time.
This prevents modified functions from being reentered before \texttt{\_DELTA} amount of time passes. 
As a result, the guarded functions cannot be called twice in the same transaction or even block, provided \texttt{\_DELTA} is greater than zero.
This has the added benefit of preventing flash loans and other interactions which may exploit a short time frame.
The duration is tied to a particular address interacting with the guarded contract, so as to prevent the guarded functionality from effectively being paused after a single interaction; other users are not restricted from accessing these functions.
One example of a timestamp-based restriction has been implemented by GMX, a popular perpetuals exchange with over \$570 million in total value locked (TVL) at the time of writing~\cite{gmx}. 
GMX does not allow a liquidity (i.e., digital assets) provider to remove liquidity until at least 15 minutes have passed since their last deposit.

This approach is not complicated, but should be used with caution.
On Ethereum, the behaviour of \texttt{block.timestamp} is well defined: it returns the timestamp in the block containing the calling transaction (though it can drift a little; see e.g., \cite{ma2019fundamentals}).
However, on related EVM-based systems which support Solidity, this may not be the case.
On the Optimism network, \texttt{block.timestamp} returns the layer two block timestamp, but on the Polygon zkEVM rollup, the value \texttt{block.timestamp} returns is not clear, and is presumably the time that a batch of transactions is processed.
As a result, on Optimism, this guard works as expected (as layer two blocks have timestamp rules that are analogous to Ethereum), but this modifier does not work on the Polygon zkEVM: many layer two blocks may have the same layer one block epoch.

\begin{figure}[bt]
    \centering
\begin{lstlisting}[language=Solidity,firstnumber=1]
abstract contract ReentrancyGuardDuration {
    uint256 private constant _DELTA = 60 seconds;
    mapping(address => uint256) public latestEntry;
    modifier nonReentrant() {
        require(latestEntry[msg.sender] + 
            _DELTA <= block.timestamp,
            "Called again too soon");
        latestEntry[msg.sender] =
            block.timestamp;
        _;
    }
}
\end{lstlisting}
\caption{Preventing access using timestamps. A function is initially not entered as the default mapping value in Solidity is \texttt{0} (line~3).
The function can only be called if the message sender has waited a sufficient amount of time with their last interaction with the contract (lines~5-7).
The \texttt{latestEntry} variable is set to the timestamp of the bock when the function is called (lines~8-9) for the caller of the function. In this case, guarded functions cannot be reentered until after \texttt{\_DELTA} time passes by each Ethereum account.}
\label{fig:timebased}

\end{figure}

%\subsection{Block Number-Based Mutex}\label{sec:block}

The guard in Figure \ref{fig:timebased} can be changed to instead consider a \texttt{\_DELTA} value in terms of block number, rather than a time period.
This can be achieved by changing setting \texttt{\_DELTA} to equal a positive integer representing the number of blocks on line~2 and changing \texttt{block.timestamp} to \texttt{block.number} on lines~6 and 9.
In this case, the modifier prevents the contract from being reentered before \texttt{\_DELTA} number of blocks pass between calls, and again, the guarded functions cannot be called twice in the same transaction or even block. 
On Ethereum there is no meaningful difference between using timestamp and block number in this way, however, this approach should be used with caution on other EVM-compatible chains.
On the Polygon zkEVM rollup, the \texttt{block.number} function returns the ``number of processable transactions'' \cite{polygontime}, which is actually the number of transactions processed on the rollup, as each block in this system contains a single transaction.
This approach will not offer the same protection on the Polygon zkEVM rollup as it might on others, like Optimism, for which the opcode is defined in terms of multi-transaction layer two blocks.

\subsubsection{Preventable Incidents}\label{sec:prev-inc-duration}

This approach may have mitigated the flash loan-based exploits discussed in Section \ref{sec:tx}.
Additionally, the second Platypus incident \cite{platypus2} was the result of repeated flash loans, sometimes minutes apart, sometimes hours apart.
Mitigating this attack would have required restricting access beyond the duration of a single transaction.

%%%%Platypus 2:
% https://cointelegraph.com/news/platypus-flash-loan-exploit-defi
% https://www.halborn.com/blog/post/explained-the-platypus-finance-hack-october-2023

\subsubsection{Trade-Offs}\label{sec:tradeoffs-duration}
These approaches trades user experience and developer experience (and a bit of gas) for likely the greatest amount of security.
A small duration lock tied to the user's deposits (see also Section \ref{sec:access}) could be used to restrict emergency functionality from being immediately callable: in reality, it is unlikely that an honest actor will trigger emergency action immediately after initiating a deposit with the transaction; likely, it will come much later.
A large duration also allows the protocol operator to pause the system if that functionality exists and they detect a problematic series of transactions.
However, the developer experience is reduced as they will need to confirm the similarity of execution semantics across all of their target blockchains.

\section{Related and Future Work}\label{sec:related}

Related work can be placed into three categories: (1)~detecting the special case of reentrant functions; (2) design and access control patterns for secure smart contracts; and (3) the analysis of DeFi protocols and attacks.
In general, the works studying the first two categories pre-date the proliferation of rollups and other mainstream layer two scaling solutions. 
In particular, none of the surveyed works discuss the possibility of different semantics on different blockchains.

Detecting and preventing malicious reentrancy has been the focus of many works.
In~\cite{DBLP:conf/ndss/RodlerLKD19}, the authors aim to automatically introduce write locks to prevent problematic reentrancy on contracts, but operate at the bytecode level. They note that as a result, it is difficult to distinguish between state updates that target an important variable (like \texttt{balance}) or the mutex lock itself.
There are several automated tools that aim to detect reentrancy (\cite{DBLP:conf/dasc/TangLB21} and \cite{8782988} provide comparisons of some of them; one example is Slither \cite{DBLP:conf/icse/FeistGG19}).
Others, like \cite{DBLP:conf/isola/DemeyerRV22}, have studied automatically refactoring code to conform to the checks-effects-interactions pattern (Section \ref{sec:checks}).
Similarly, the works of \cite{DBLP:conf/uss/RodlerLKD21, DBLP:conf/wcre/ZhangMLLNG20} introduce bytecode rewriting tools that can rewrite smart contract bytecode to avoid many classes of bugs while keeping the intended functionality of the smart contract.

Access restriction to dApps has also been the subject of academic study.
%Access restriction via hash function can be used along with its pre-image for locking and unlocking the contract for entry~\cite{DBLP:conf/icccnt/SP23}.
Instead of preventing access to functions for a specified duration, one can restrict by requiring a pre-image for a hash to be supplied during the call \cite{DBLP:conf/icccnt/SP23}.
In~\cite{DBLP:journals/fcomp/AlkhalifahNWK21}, a method to check that a pre- and post-condition for a function call is necessary for the function to be considered non-reentrant. 
They have a specific use case in mind where the balances of the contract are as expected before and after the call; in a sense, this sidesteps the reentrancy mitigation, opting instead to check the results of a potentially catastrophic reentrant call.
Cecchetti et al.~\cite{DBLP:conf/sp/CecchettiYNM21} present a security type system that provably enforces secure information flow, which can prevent unintended system behaviour, but their approach requires additional developer overhead.
In particular, they require annotations on code and additional tooling, rather than the Solidity-only approaches we presented.
While their approach may be more powerful than our techniques, it is also likely less practical if the related tooling fails to mature alongside Solidity and the myriad of EVM-compatible chains.

%DeFi Protocols
There are fewer works that highlight the impact of the (lack of) these design patterns systematically.
DeFi incidents are evaluated in a ``common reference frame'' to distinguish between reports provided by academia and practitioners in~\cite{DBLP:conf/sp/ZhouXECWWQWSG23}. 
While they note that incentives on layer two protocols may differ (affecting the use of these protocols), they do not mention that in fact the underlying execution layer may affect redeployment of these protocols. 
They also do not dive into the particulars of any specific incidents, unlike in this work.

In~\cite{10.1145/3532857}, the authors model the interaction between various DeFi protocols.
One direction for future work is to refine this model to include how many times a protocol must call a function on another one during an non-malicious transaction. 
This would confirm the applicability of the modifiers presented in this work to off-the-shelf protocols and suggest necessary requirements for future versions of them.

A more immediate research direction is to study DeFi protocols on different EVM-based blockchains and determine if any additional risk has been introduced. 
In particular, a framework to evaluate the fitness of an existing protocol for another blockchain with a different implementation of the EVM should be established and maintained by the community; websites like \href{https://www.rollup.codes/}{RollupCodes}\footnote{\url{https://www.rollup.codes/}} and \href{https://l2beat.com/}{L2Beat}\footnote{\url{https://l2beat.com/}} may serve as good starting points.
It will also be important to revisit the solutions presented in this work when EIP-1153 \cite{ERC-1153} is accepted.
EIP-1153 introduce a way to store variables that are discarded at the end of a transaction, and could be used for the variables within the discussed function modifiers.
% Would be helpful to study the average deposit/lockup time for DeFi to justify approaches that force a delay

\section{Conclusion}\label{sec:conclusion}

We have presented some methods to prevent and mitigate the effect of malicious repeated interactions with smart contracts.
Though they each have their own trade-offs, these methods could have been used to prevent \$136M worth of DeFi losses in the last year alone.
We have highlighted the risk of using the same mitigation methods across various EVM based blockchains as the semantics of various built-in functions may change.
Almost every exploit could have been prevented by merely disallowing reentrancy altogether, a problem thought to be well understood by the community.
However, as we have shown by illustrating a read-only reentrancy attack, this topic requires further study.
We are drawing attention to related concerns (like different execution semantics) now, in an attempt to prevent future catastrophes.
\paragraph*{Acknowledgements.}

The authors would like to thank their colleagues, especially Mathias Hall-Andersen, for their insights and fruitful inputs that were incorporated in this paper.

\bibliographystyle{unsrt}
\bibliography{references}

\end{document}